\documentclass[11pt]{article}
\usepackage{blois,epsfig}

\def\Journal#1#2#3#4{{#1} {\bf #2}, #3 (#4)}

\def\PLB{{\em Phys. Lett.}  B}
\def\PRL{\em Phys. Rev. Lett.}
\def\PRD{{\em Phys. Rev.} D}

\def\NP{\em Nucl. Phys.}

\def\mco{\multicolumn}

\def\ra{\rightarrow}

\def\ko{K^0}

\def\be{\begin{equation}}
\def\ee{\end{equation}}
\def\bea{\begin{eqnarray}}
\def\eea{\end{eqnarray}}

\begin{document}
\vspace*{4cm}
\title{NEUTRINOS POLARIMETRY IN MUON DECAY}

\author{ S.J. CIECHANOWICZ, W. SOBK\'OW}

\address{Institute of Theoretical Physics,  University of Wroc\l{}aw, Pl. M. Borna 9,
\\PL-50-204~Wroc\l{}aw, Poland}

\author{M. MISIASZEK}

\address{ M. Smoluchowski Institute of Physics, Jagiellonian University, ul. Reymonta 4,\\
PL-30-059 Krak\'ow, Poland}

\maketitle
\abstracts{We consider massive Dirac electron antineutrino to be observed in the muon decay, $\mu^{-} \rightarrow e^- +\overline{\nu}_{e} + \nu_{\mu}$, at rest in the LAB reference frame. A computation of $\overline{\nu}_{e}$ differential direct spectral shape and angular distribution was performed. This has resulted in transverse component of the neutrino polarization emerging from the nonstandard $S $, $T $ sector of the Lorentz structure in the Fermi leptonic charged weak interaction. The calculation is based on empirical values of the coupling parameters.}

\section{Introduction}\label{sec:intro}

Main purpose of our work is to show that in case of stopped muon decay, if measuring the emission of the massive electron antineutrino, nonstandard couplings for the right-handed neutrino could be tested, see also \cite{ft}.

In general, the importance of the question about the nature of the leptonic weak couplings
is growing in the context of finite masses of the neutrino family and discovering the processes changing the lepton charge numbers. The muon rationale then is twofold: precision testing limits in standard weak interactions, and pursuing new physics in nonstandard sector.

Since, Cecilia Jarlskog analysis \cite{ja} of muon decay coupling constants in the Lorentz structure general frame, it was recognized that decay parameters are mutually related and the data give ambiguous results on the coupling constants.

In this paper we have used the notations presented in paper \cite{am}. For calculations and results presentation we took the advantage of chirality representation formalism for the lepton states in the decay Hamiltonian.

In Sec. 2 of this work, we present the formula of the differential probability in muon decay, and give the number upper limit on polarization transverse component of the electron antineutrino. The paper is concluded in Sec. 3.

\section{Electron Antineutrino Polarization}\label{sec:beam}

\begin{table}[t]
\caption{\label{tab:le1} Current limits on the non-standard coupling factors. }
\vspace{0.4cm}
\begin{center}
\begin{tabular}{*{12}{l}}
\hline
$|g_{LL}^V|>0.960$   & $|g_{LR}^V|<0.036$ & $|g_{RL}^V|<0.104$ & $|g_{RR}^V|<0.034  $ \\
$|g_{LL}^S|<0.550$   & $|g_{LR}^S|<0.088$ & $|g_{RL}^S|<0.417$ & $|g_{RR}^S|<0.067  $ \\
$|g_{LL}^T|\equiv 0$ & $|g_{LR}^T|<0.025$ & $|g_{RL}^T|<0.104$ & $|g_{RR}^T|\equiv 0$ \\
\hline
\end{tabular}
\end{center}
\end{table}

Weak forces dynamics with parity violation, we work on here, is the four fermion theory of contact interactions. Now then, our only extension of Standard Model consists in adding $R $-handed neutrino that interacts via scalar and tensor forces instead of vector one. Also, neutrinos of definite $L $- and $R $-chirality are massive Dirac states in the astrophysical limits range of several $eV $.

\begin{figure}
\vskip 0.5cm
\begin{center}
\psfig{figure=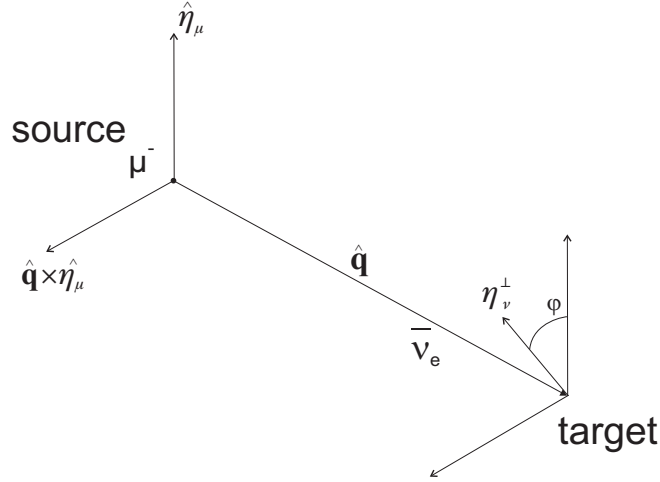,height=2.5in}
\end{center}
\caption{Kinematics for stopped muon decay in the LAB reference system. Vectors $\mbox{\boldmath ${\hat \eta}_{\mu}$}$, muon polarization direction and $\hat{\bf q}$, neutrino momentum direction span the neutrino production plane. $\mbox{\boldmath ${\hat\eta}_{\overline{\nu}}^{\perp}$} $ is the transverse polarization of the neutrino. The azimuthal angle $\phi $ is the $CP $-violating phase.
}\label{fig:ure1}
\end{figure}

The transition matrix element of the polarized muon decay is, see in W. Fetscher and H.-J. Gerber \cite{am,fe,ha}:
\bea
M_{\mu^{-}} & = &
\frac{G_{F}}{\sqrt{2}}\{g_{LL}^{V}(\overline{u}_{e}\gamma_{\alpha}(1-\gamma_5)v_{\nu_{e}})
(\overline{u}_{\nu_{\mu}} \gamma^{\alpha}(1 - \gamma_{5})u_{\mu})             \nonumber \\
& &  \mbox{} + g_{LR}^{S}(\overline{u}_{e} (1+\gamma_5)v_{\nu_{e}})(\overline{u}_{\nu_{\mu}}
(1 + \gamma_{5})u_{\mu})                                                      \nonumber \\
& & \mbox{} + \frac{g_{LR}^{T}}{2}(\overline{u}_{e} \sigma_{\alpha\beta}(1+\gamma_5)v_{\nu_{e}})
(\overline{u}_{\nu_{\mu}}\sigma^{\alpha\beta}(1 + \gamma_{5})u_{\mu})\},
\label{eq:amplit}
\eea
where $g^{V}_{LL}$, $g^{S}_{LR}$ and $g_{LR}^T$ are the dimensionless coupling factors with superscripts $V, S, T $  indicating the type of weak interaction while subscript indexes $L, R$ both amount to the chiralities for either daughter electron antineutrino or initial muon, the number values \cite{am} we present in Table \ref{tab:le1}.
Because, we allow for the non-conservation of the combined symmetry CP,
the coupling constants $g_{LL}^V, g_{LR}^S $ and $g_{LR}^T $ are complex numbers.
In the SM, $g_{LL}^{V}=1$ and $g_{LR}^S $, $g_{LR}^T \equiv 0 $.

The outgoing electron antineutrino flux is a mixture of the left-chirality antineutrinos produced in the $g_{LL}^V$ weak interaction and the right-chirality ones produced in the  $g_{LR}^S$ and the $g_{LR}^T$ weak interactions.
In the limit of vanishing antineutrino mass, the left-chirality antineutrino has positive helicity, while the right-chirality one has negative helicity, see \cite{fe}.
The muon neutrino is always left-chirality both for the $g_{LL}^V$  and the $g_{LR}^S$, $g_{LR}^T$ couplings (muon neutrino has negative helicity, when $m_{\nu_\mu}\rightarrow 0)$.
It means that the neutrino chirality is the same as the associated charged lepton for the $V $ interaction, and opposite for the $S, T $ interactions.
There are only two non-zero interferences between the standard coupling $g_{LL}^V$ and exotic  couplings, i. e. $g_{LR}^S$ and $g_{LR}^T$.

With some Dirac matrices algebra \cite{gr}, the differential probability for the energy and angular distribution of the electron antineutrinos in the polarized muon decay at rest is given as below:
\bea
\label{eq:beam}
 \frac{d^2 \Gamma}{ dy d\Omega_\nu}
& = &
   \frac{d^2 \Gamma_V}{ dy d\Omega_\nu}
+  \frac{d^2 \Gamma_{S + T}}{dy d\Omega_\nu}
+  \frac{d^2 \Gamma_{VS + VT}}{dy d\Omega_\nu} ,
\\
   \frac{d^2 \Gamma_{V}}{ dy d\Omega_\nu}
 & = & \frac{G_{F}^2 m_{\mu}^5 }{128\pi^4}\Bigg\{|g_{LL}^{V}|^2 y^2 (1 - y)
(1+\mbox{\boldmath $\hat{\eta}_{\overline{\nu}}$}\cdot\hat{\bf q})
(1+P_{\mu}\mbox{\boldmath $\hat{\eta}_{\mu}$}\cdot\hat{\bf q})\Bigg\},
\label{eq:V}
\\
   \frac{d^2 \Gamma_{S + T}}{ dy d\Omega_\nu}
& = &
   \frac{G_{F}^2 m_{\mu}^5}{3072\pi^4}(1-\mbox{\boldmath
$\hat{\eta}_{\overline{\nu}}$}\cdot\hat{\bf q}) y^2\Bigg\{|g_{LR}^{S}|^2
\bigg[(3 - 2y) - (1 - 2y)P_{\mu}\mbox{\boldmath
$\hat{\eta}_{\mu}$}\cdot\hat{\bf q}\bigg] \nonumber
\label{eq:S+T}
\\
& & \mbox{} + 4 \left|g_{LR}^{T} \right|^2\bigg[(15 - 14y) -
(13 - 14y)P_{\mu}\mbox{\boldmath $\hat{\eta}_{\mu}$}\cdot\hat{\bf q} \bigg]\Bigg\},
\\
\frac{d^2 \Gamma_{VS + VT}}{dy d\Omega_\nu} & = &
\frac{G_{F}^2 m_{\mu}^5}{256\pi^4} y^2(1 - y)P_{\mu} \Bigg\{Re\big[ g_{LL}^V( g_{LR}^{S}
- 6g_{LR}^{T} )^* \big](\mbox{\boldmath $\eta_{\overline{\nu}}^{\perp}$}\cdot \mbox{\boldmath $\hat{\eta}_{\mu}$}) \nonumber
\\
& & + Im\big[ g_{LL}^V( g_{LR}^{S}- 6g_{LR}^{T} )^* \big]
\mbox{\boldmath $\eta_{\overline{\nu}}^{\perp}$}\cdot({\bf \hat{q}} \times
\mbox{\boldmath $\hat{\eta}_{\mu}$})\Bigg\},
\label{eq:VS+VT}
\eea \noindent
where the notations are:
variable $y=2E_{\nu}/m_{\mu}$ is the reduced antineutrino energy with the muon mass $m_\mu$, it varies from $0 $ to $1$, $d\Omega_\nu$ is the solid angle differential for $\overline{\nu}_e$ linear momentum unit vector $\hat{\bf q}$, $G_{F} $ is the universal Fermi coupling constant \cite{am}, the unit polarization vector of the electron antineutrino $\overline{\nu}_e$ in its rest reference is denoted by
$\mbox{\boldmath $\hat{\eta}_{\overline{\nu}}$}$, the scalar product $(\mbox{\boldmath$\hat{\eta}_{\overline{\nu}}$}\cdot\hat{\bf q})$
is a value of the longitudinal polarization component of
$\mbox{\boldmath $\hat{\eta}_{\overline{\nu}}$}$, and $\mbox{\boldmath $\hat{\eta}_{\overline{\nu}}^{\perp}$}$ stands for the vector of the polarization transverse component against the vector $\hat{\bf q}$, Fig. \ref{fig:ure1}.
The symbols $P_{\mu} $ and $\mbox{\boldmath $\hat{\eta}_{\mu}$}$, respectively, are the value of the muon spin polarization and the unit vector of its direction.
Nota bene, precise measurement on the number value of $P_{\mu} $ would be necessary.

After normalizing the coupling factors, the inverse of the muon lifetime formula has been obtained \cite{fe}:

\be
\tau^{-1}_{\mu}  = \frac{G_{F}^{2} m_\mu^5}{192 \pi^{3}}\left(|g_{LL}^{V}|^2 +
\frac{1}{4}|g_{LR}^{S}|^2 + 3|g_{LR}^{T}|^2 \right),
\ee

where:

\be
|g_{LL}^{V}|^2 + \frac{1}{4}|g_{LR}^{S}|^2 + 3|g_{LR}^{T}|^2 =1.
\ee

Following the definitions of \cite{fs}, and using the data from \cite{am,ha}, we have calculated the probability $Q_{L}^{\overline{\nu}}$ for the antineutrino $\overline{\nu}_{e}$ to be left-chirality particle:

\bea \label{eq:trlo}
Q_{L}^{\overline{\nu}} &=& 1 - \frac{1}{4}|g_{LR}^S|^{2} - 3 |g_{LR}^T|^{2}\geq 0.996,
\\
\mbox{\boldmath $\hat{\eta}_{\overline{\nu}}$}\cdot\hat{\bf q} &=& 2
Q_{L}^{\overline{\nu}} -1 \geq 0.992,
\;
|\mbox{\boldmath $\eta_{\overline{\nu} }^{\perp}$}|
= 2\sqrt{Q_{L}^{\overline{\nu}}(1-Q_{L}^{\overline{\nu}})}  \leq 0.128,
\eea
finally, we obtain the lower limit on the longitudinal antineutrino polarization, $(\mbox{\boldmath $\hat{\eta}_{\overline{\nu}}$}\cdot\hat{\bf q})  $, and upper bound on the magnitude of the polarization transverse component, $|\mbox{\boldmath $\eta_{\overline{\nu} }^{\perp}$}|  $.
This example illustrates the transverse component of the neutrino polarization in muon decay being sensitive to non-standard coupling parameters.

Eq. \ref{eq:VS+VT} is the term that includes interferences between the standard
$g_{LL}^{V}$ and exotic, non-standard $g_{LR}^S, g_{LR}^T$ coupling parameters, so, contrary to Eq. \ref{eq:S+T}, this term is linear in the exotic couplings. Moreover, it can be seen that neutrino transverse component observable is divided into $CP $-conserving and $CP $-breaking parts,
$(\mbox{\boldmath $\eta_{\overline{\nu}}^{\perp}$}\cdot \mbox{\boldmath $\hat{\eta}_{\mu}$})  $ and
$\mbox{\boldmath $\eta_{\overline{\nu}}^{\perp}$}\cdot({\bf \hat{q}} \times
\mbox{\boldmath $\hat{\eta}_{\mu}$})  $, respectively.

\section{Discussion}

In this paper, a proposition for the polarized electron antineutrino spectrum measurement in polarized muon decay is considered. Using the experimental limitations of the coupling constants within the framework of Lorentz structure, the spectral shape and electron-antineutrino angular correlation has been computed.

Existing data yet admit small deviations from  $V - A $ coupling type leaving space for the interacting right-handed neutrino in addition to the standard left-handed one.
As a result, polarization vector of the electron-antineutrino may acquire a transverse component.

Concluding on neutrino polarimetry, we shall outline a feasibility to detect neutrino spin polarization in laboratory. At first, we refer KARMEN experiment \cite{ka}, the high resolution neutrino spectrometer to be replaced with the neutrino polarimeter. With this aim, for the neutrino source the polarized muons facility could be installed, while in the neutrino detector, the polarized finite spin nuclei could be used as a target for the neutrinos.
Second, according to measurement propositions for muon decays; it was scattering neutrino beam on polarized electron target proposed to search the neutrino magnetic moment \cite{ra}, probe the neutrino beam flavour composition \cite{mi}, and look for the right handed neutrinos \cite{ci}.

Now, in a separate article, we shall continue direct research of the neutrino polarization in muon decay mainly to analyze the potentiality of electron antineutrino elastic collisions with the unpolarized detector electrons to measure the right handed neutrino couplings.

\section*{Acknowledgments}

The author, S. C., is deeply indebted to Professor Jean Tr\^an Thanh V\^an for funding the conference expenses and all of the organizers for their kind assistance.


\section*{References}

\begin{thebibliography}{99}

\bibitem{ft} W. Fetscher, \Journal{\PRL}{69}{2758}{1992}.

\bibitem{ja} C. Jarlskog, \Journal{\NP}{75}{659}{1966}.

\bibitem{am} C. Amsler {\it et al.}, \Journal{\PLB}{66}{1}{2008}.

\bibitem{fe} W. Fetscher, \Journal{\PRD}{49}{5945}{1994}. 

\bibitem{fs} W. Fetscher {\it et al.}, \Journal{\PLB}{173}{102}{1986}.

\bibitem{ha} K. Hagiwara {\it et al.}, (Particle Data Group), \Journal{\PRD}{66}{010001}{1992}.

\bibitem{gr} W. Greiner and B Mueller, Gauge Theory of Weak Interactions, Springer, 2000.

\bibitem{ka} B. Armbruster {\it et al.}, \Journal{\PRL}{81}{520}{1998}.

\bibitem{ra} T. I. Rashba {\it et al.}, \Journal{\PLB}{541}{151}{2002}.

\bibitem{mi} P. Minkowski {\it et al.}, \Journal{\PLB}{479}{218}{2000}.

\bibitem{ci} S. Ciechanowicz {\it et al.}, \Journal{\PRD}{71}{093006}{2005}


\end{thebibliography}
\end{document}